\documentclass[twocolumn,preprintnumbers,amsmath,amssymb]{revtex4}
\def
\<{\langle}\def
\<{\rangle}
\usepackage{graphicx}
\usepackage{bm}

\begin{document}


\title{Comments on the Discrete Variable Representation}

\author{Barry I. Schneider}
\email{bschneid@nsf.gov}
\affiliation{Physics Division, National Science Foundation, Arlington,
Virginia 22230 and Electron and Optical Physics Division, National Institute
of Standards and Technology, Gaithersburg, MD 20899} 
\author{Nicolai Nygaard}
\email{nicolai.nygaard@nist.gov}
\affiliation{Chemical Physics Program,
             University of Maryland, College Park, MD 20742-2431 and
             Electron and Optical Physics Division, National
             Institute of Standards and Technology, Gaithersburg, MD 20899}        

\date{\today}

\begin{abstract}
We discuss the application of the Discrete Variable Representation to
Schr\"odinger problems which involve singular Hamiltonians. Unlike 
recent authors who invoke transformations to rid the eigenvalue equation
of singularities at the cost of added complexity, we show that an
approach based solely on an orthogonal polynomial basis is adequate,
provided the Gauss-Lobatto or Gauss-Radau quadrature rule is used.  This ensures
that the mesh contains the singular points and by simply discarding the DVR functions
corresponding to those points, all matrix elements become well-behaved, the 
boundary conditions are satisfied and the calculation is rapidly convergent.  
The accuracy of the method is demonstrated by applying it to 
the hydrogen atom. We emphasize that the method is equally capable 
of describing bound states and continuum solutions. 

\end{abstract}

\maketitle
\section{Introduction}
The Discrete Variable Representation (DVR) \cite{Dick&Cert,Light,Wyatt,
Mucker, Baye&Heenen,Schn&Nygaar} is one of the most effective and 
widely used methods for discretizing the Schr{\"{o}}dinger equation.  
In its most elemental form, it has the virtues of maintaining the 
locality of operators which are local in space, and the rapid 
convergence of a spectral method.  In addition, for multi-dimensional problems it leads to a sparse 
matrix representation of the Hamiltonian, which may be used quite 
effectively when coupled to iterative techniques designed to solve
large sets of linear equations or to extract the lowest eigenvalues of 
large matrices.  A recent variant of the method, which combines the DVR 
with a finite element method \cite{Res&McC}, has been used to solve 
one of the most intractable problems in atomic scattering theory, the 
impact ionization of the hydrogen atom.  Lately, the technique has 
been combined with the Arnoldi/Lanczos approach to produce an
extremely efficient method for the solution of the 
time-dependent Schr{\"{o}}dinger equation \cite{SchCol&Fed}.

The purpose of this note is to correct some misconceptions concerning the
application of the method to problems involving singular potentials.  
These issues appear to arise when it is apparent that the boundary 
conditions satisfied by the solution to the Schr{\"{o}}dinger 
equation should not lead to any numerical difficulties.  A number of 
authors~\cite{Baye&Heenen,VinckeMalegat&Baye,Malegat&Vincke,Malegat}
have provided "remedies" to remove the singularities and to  
transform the original Schr{\"{o}}dinger equation in to a more tractable 
and rapidly converging form.
Unfortunately, these transformations often destroy the natural symmetry 
of the original equations and lead to more complex algebraic solution 
methods than is really necessary.
Here we present an alternative approach, which addresses the problem more transparently leading to a simpler numerical procedure with no loss of accuracy.
Section~\ref{DVRsec} is a summary of the key elements of the DVR method, and in section~\ref{SingHam} we present our approach for applying this methodology to singular Hamiltonians. We end in section~\ref{Conclusions} with a brief conclusion.

\section{Discrete Variable Representation}
\label{DVRsec}
Since the DVR has been discussed extensively \cite{Dick&Cert,Light,Wyatt,
Mucker, Baye&Heenen,Schn&Nygaar} in the literature, we provide only the 
essentials here.  A DVR exists when there is both a spectral basis of 
$N$ functions, $\phi_{i}(x)$, orthogonal over a range [a,b] with weightfunction $w(x)$
\begin{equation}
\int_a^b w(x) \phi^*_n(x) \phi_m(x) dx = \delta_{m,n},
\end{equation} 
and an associated quadrature rule with $N$ points, 
$x_{i}$ and weights, $w_{i}$ which enable a set of coordinate eigenfunctions 
$u_{i}(x)$ to be defined  with the following properties,
\begin{subequations}
\label{coordfn}
\begin{eqnarray}
 u_{i}(x) &=& \sqrt{w(x)} \sum_{n=0}^{N-1} c_{n} \phi_{n}(x), \\  
 u_{i}(x_{j}) &=& \sqrt{\frac{ w(x_i) } {w_{i}}} \delta_{i,j}.  
\end{eqnarray}
\end{subequations}
Using the quadrature rule to evaluate $c_{n}$ gives,
\begin{subequations}
\label{DVR}
\begin{eqnarray}
\label{dvr1}
c_{n} 
      &=& \int_{a}^{b} { \sqrt{w(x)} \phi^*_{n}(x) u_{i}(x) dx } \nonumber \\
      &=& \sum_{k=1}^N  w_k \phi^*_n(x_k) \frac{u_i(x_k)}{\sqrt{w(x_k)}} \nonumber \\
      &=& \sqrt{w_{i}} \phi^*_{n}(x_{i}), \\
\label{dvr2}
 u_{i}(x) &=& \sqrt{w_{i} w(x)}  \sum_{n=0}^{N-1} \phi^*_{n}(x_{i}) 
                               \phi_{n}(x)  \\
\label{dvr3}
     \langle u_{i} \mid x \mid u_{j} \rangle &=&
\delta_{i,j} x_{i}.
\end{eqnarray}
\end{subequations}
There are two important features to note.  First, the coordinate 
eigenfunctions are defined as continuous functions of the spectral
basis. When this basis is polynomial the sum in Eq.(\ref{dvr2}) can be
carried out exactly, and the coordinate eigenfunctions can be
expressed as
\begin{equation}
u_{i}(x) = \sqrt{\frac{w(x)}{w_{i}}} 
              { \prod_{k=1}^{N} }{}' \, \frac{x-x_{k}}{x_{i}-x_{k}},
\end{equation}
the Lagrange interpolating functions at the 
quadrature points.   With either representation, they may be easily 
differentiated analytically.  Second, the expansion coefficients, 
$c_{n}$, are computed using the quadrature rule.  Implicit in
using the quadrature rule for the evaluation of $c_{n}$ is that the
result is accurate.  This is not guaranteed except for certain cases.  For
example, when $\phi_{i}(x)$ is one of the classical orthogonal functions,
there is an associated Gauss quadrature \cite{Kopal} which guarantees that
Eq.(\ref{DVR}) is exact when the integrand is a polynomial of degree 
$(2N-1)$ or less.  There are other examples such as particle-in-a-box or 
Fourier functions, which are not polynomials, but which can be shown to 
exactly satisfy Eq.(\ref{DVR}) with an appropriately chosen quadrature
rule.  In all of these cases there exists a unitary transformation 
between the original spectral and coordinate basis.  Since the 
coordinate functions diagonalize the coordinate operator, any function 
of the coordinates is also diagonal.  This is very convenient for 
actual calculations and gives the DVR calculation many of the 
desirable properties of a grid based method with few of the
disadvantages.  It should also be noted that matrix elements of the 
kinetic energy operator while not diagonal in the coordinate 
basis may be evaluated simply and exactly using the quadrature rule or 
analytically.  Since the kinetic energy part of the Hamiltonian matrix 
is a separable sum over particle and coordinate variables, a product DVR 
basis leads to a sparse representation. 
When the interval [a,b], is infinite or semi-infinite, the weight 
function, $w(x)$, insures that the wavefunction will decay properly at 
large distances.  For finite intervals, boundary conditions may be 
enforced by requiring that the wavefunction or its derivative behave 
correctly at the left and/or right boundary.

There is a simple, but quite useful generalization of Gauss quadratures 
that will be needed in what follows.  It is possible to specify in 
advance that some of the points are fixed.  When these points are either 
or both of the endpoints of a finite interval, the quadrature rule is 
termed a Gauss-Radau or Gauss-Lobatto quadrature, respectively.
The remaining Gauss points may be determined by a simple modification
of the original procedure~\cite{Kopal}.  Since one or two points are now fixed, the  
quadrature is of lower accuracy than the full Gauss quadrature, but the 
great advantage of being able to satisfy specific boundary 
conditions at the endpoints, far outweighs this disadvantage.

\section{Singular Hamiltonians}
\label{SingHam}
Consider the radial Schr\"odinger equation,
\begin{equation}
\label{radial wave equation}
\big [ - \frac {1}{2} \frac{d^2}{dr^2} + 
          \frac{l(l+1)}{2r^2} + v(r) - E \big ] \psi(r)  = 0
\end{equation}
where we assume that $v(r)$ vanishes for large $r$ and is singular at the origin. The radial function 
satisfies the boundary condition $\psi(0)=0$,  
and either exponentially decays or oscillates 
for large $r$.  Here we will offer two alternative approaches to
solving Eq.(\ref{radial wave equation})  To motivate the discussion,
recall that Baye and Heenen \cite{Baye&Heenen} suggest 
that for the case of exponentially decaying boundary conditions,
one very natural choice for the spectral functions is,
\begin{equation}
\phi_{n}(r) = r^{l+1} \exp(-r/2) L_{n}^{2l+2}(r)
\end{equation} 
where $L_{n}^{\alpha}(r)$ are the generalized Laguerre polynomials.
When this basis is used for the coulomb potential, the results are quite disappointing.  
The relative error in the ground state energy with ten basis functions is about $5\cdot 10^{-3}$. 
This appears to be simply related to the choice of $r^2  \exp(-r)$ as the
weight function.  While this choice does result in a set of coordinate
functions which satisfy both boundary conditions, it gives rise to a
potential energy matrix element which does not behave as a polynomial
times the weight function.  In fact, the integrand has terms which behave
as inverse powers of $r$.  Vincke, Malegat and Baye \cite{VinckeMalegat&Baye}
propose a simple procedure to remedy the problem.  They regularize the
problem by multiplying the Schr\"odinger equation by $\rho(r)$, 
where, $\rho(r)$ is chosen so that, $\rho(r) v(r) = $ constant as $r=0$. 
Using, for example, $\rho(r)=r^2$, leads to a generalized eigenvalue 
problem with a modified kinetic energy matrix based on Laguerre polynomials
with $\alpha = 0$.  Here we suggest a more direct attack.  First, we do 
not transform the Schr\"odinger equation.  We use the Laguerre polynomials 
with $\alpha = 0$, that is, with a weight function, $ \exp(-r)$, but choose 
the points and weights of the quadrature by the Gauss-Radau rule with $r=0$ as 
the fixed point.  The set of resulting DVR functions all satisfy the 
boundary conditions at infinity and due to the Kronecker delta function 
property~(\ref{coordfn}b) all but the first DVR function also satisfy the 
boundary condition at the origin, that is, they lead off as $r$.  The first basis
function is then simply dropped from the expansion.  The resulting 
matrix elements of the Hamiltonian are all exactly integrated by the 
quadrature rule and quite well behaved. 

We have applied our method to the spectrum of the hydrogen atom. In Fig.~1 we show the relative error $\varepsilon$, on the first ten eigenstates with $l=0$ for various basis set sizes. For comparison we also plot the results obtained when using the regularized mesh technique of Vincke {\emph{et. al.}} (with scaling factor $h=1$, see~\cite{VinckeMalegat&Baye}). In addition to its greater simplicity the accuracy of our method is equal or superior to that of the regularized mesh technique. Moreover, since all basis functions vanish at the origin, our method works equally well for finite values of the angular momentum, as long as the wavefunction is well localized within the interval. 
 
\begin{figure}
\label{Lagerrors}
\begin{center}
\includegraphics[scale=0.45]{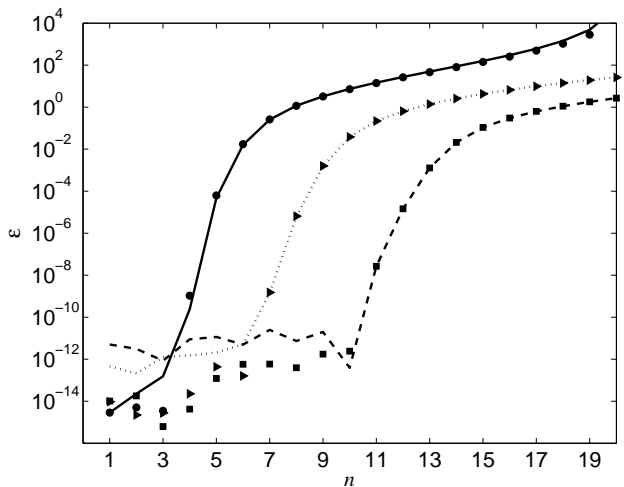}
\caption{Relative error on the first ten $l=0$ eigenstates of hydrogen using a Gauss-Laguerre basis with no scaling (h=1). The points indicate the results obtained using the method of this paper for $N=20$ ($\bullet$), $N=50$ ($\blacktriangleright$), and $N=100$ ($\blacksquare$). The lines represent the relative error obtained using the regularized Lagrange mesh method of Vincke {\emph{et. al.}}~\cite{VinckeMalegat&Baye} for $N=20$ (solid), $N=50$ (dots), and $N=100$ (dashed).}  
\end{center}   
\end{figure}

A second approach, which works for both the bound and continuous spectrum,
places the system in a large box of radius, $r=a$.  
The DVR basis is defined using the Gauss-Legendre-Lobatto quadrature rule.  
By ensuring that the two endpoints are part of the quadrature, it becomes 
trivial to satisfy the boundary conditions.  Dropping the DVR function 
at the origin, guarantees that the solution will vanish at $r=0$.  
If the DVR function at the last point is dropped, the solution will go 
to zero at $r=a$ and simulate exponentially decaying solutions.  By
retaining the DVR function at the last point and adding a Bloch 
operator,
\begin{equation}
L =   \frac { {\hbar}^2 }{2M} [ \delta(x-a) \frac{d}{dx} ]
\end{equation}
to the Hamiltonian, it is possible to deal with non-fixed 
node boundary conditions at the right endpoint and simulate 
scattering boundary conditions.
For long range potentials, such as the coulomb potential, it is 
necessary to make sure that the results are not box size dependent.
Stated differently, one must examine the 
convergence of the eigenvalues with respect to basis set and box size.
This is clearly evidenced in Tables I-III where one sees convergence 
to eigenvalues of the truncated coulomb potential when the size of
the box is too small.  By systematically increasing the box size and
the basis, it is possible to obtain the eigenvalues to arbitrary
accuracy.    

\section{Conclusions}
\label{Conclusions}
Previous researchers have developed DVR techniques that require special treatment of singular
potentials or non-polynomial based quadratures.  Here we have demonstrated that a judicious
use of the orthogonal polynomial approach, using the Gauss-Lobatto quadrature rule, avoids
the need to transform the Schr{\"{o}}dinger equation into a form which is numerically less
tractable.  In addition, the method is applicable to all types of boundary conditions and is able
to treat the bound and continuous spectrum on equal footing.  As a final note, using 
the finite element DVR, enables one to treat singularities or even 
discontinuities {\cite{Res&McC} at interior points, if they are known in advance, 
by chosing the boundaries of the elements at those points.

%


\begin{table*}
\caption{s-Wave Eigenvalues of Hydrogen Atom in Legendre Basis; R=50au }
\begin{ruledtabular}
\begin{tabular}{ccccc}

$n$ &        $N=10$    &    $N=20$     &     $N=40$     &      Exact    \\
 1  &     -0.39428839  & -0.49997882   &  -0.50000000   &   -0.50000000 \\
 2  &     -0.11142228  & -0.12500000   &  -0.12500000   &   -0.12500000 \\
 3  &     -0.05165408  & -0.05555555   &  -0.05555555   &   -0.05555555 \\
 4  &     -0.02957707  & -0.03120434   &  -0.03120434   &   -0.03120434 \\
 5  &     -0.01651543  & -0.01786476   &  -0.01786476   &   -0.01786476 \\
 6  &     -0.00060937  & -0.00226590   &  -0.00226590   &   -0.00226590 \\
\end{tabular}
\end{ruledtabular}
\label{Leg1}
\end{table*}

\begin{table*}
\caption{s-Wave Eigenvalues of Hydrogen Atom in Legendre Basis; R=100au }
\begin{ruledtabular}
\begin{tabular}{ccccc}

$n$ &      $N=20$      &     $N=40$    &     $N=50$     &     Exact      \\
 1  &   -0.48882286    &  -0.50000000  &  -0.50000000   &  -0.50000000   \\
 2  &   -0.12481146    &  -0.12500000  &  -0.12500000   &  -0.12500000   \\
 3  &   -0.05554641    &  -0.05555556  &  -0.05555556   &  -0.05555556   \\
 4  &   -0.03124909    &  -0.03125000  &  -0.03125000   &  -0.03125000   \\
 5  &   -0.01999983    &  -0.01999997  &  -0.01999997   &  -0.01999997   \\
 6  &   -0.00959636    &  -0.01386848  &  -0.01386848   &  -0.01386848   \\

\end{tabular}
\end{ruledtabular}
\label{Leg2}
\end{table*}

\begin{table*}

\caption{s-Wave Eigenvalues of Hydrogen Atom in Legendre Basis; R=200au).}
\begin{ruledtabular}
\begin{tabular}{cccc}
$n$ &      $N=40$    &     $N=50$   &     Exact       \\
 1  &   -0.49997974  &  -0.49999999 &  -0.50000000    \\   
 2  &   -0.12500000  &  -0.12500000 &  -0.12500000    \\
 3  &   -0.05555556  &  -0.05555556 &  -0.05555556    \\
 4  &   -0.03125000  &  -0.03125000 &  -0.03125000    \\
 5  &   -0.0200000   &  -0.02000000 &  -0.02000000    \\
 6  &   -0.01388889  &  -0.01388889 &  -0.01388889    \\
 7  &   -0.01020408  &  -0.01020408 &  -0.01020408    \\
 8  &   -0.00781238  &  -0.00781238 &  -0.00781238    \\

\end{tabular}
\end{ruledtabular}
\label{Leg3}
\end{table*}


\begin{thebibliography}{99}

\bibitem{Dick&Cert} A. S. Dickenson and P. R. Certain, J. Chem. Phys.
{\bf{49}}, 1515 (1965).
\bibitem{Light} J. C. Light, I. P. Hamilton and J. V. Lill, J. Chem. Phys.
                {\bf 82}, 1400 (1985); S. E. Choi and J. C. Light, J. Chem.
                Phys. {\bf 90}, 2593 (1989)
\bibitem{Baye&Heenen} D. Baye and P.-H. Heenen, J. Phys. A:Math. Gen. 
                      {\bf 19 }, 2041(1986)
\bibitem{Wyatt} D. E. Manolopoulos and R. E. Wyatt, 
Chem. Phys. Letts. {\bf152}, 23 (1988)
\bibitem{Mucker} J. T. Muckerman, Chem. Phys. Letts. {\bf 173}, 200 (1990);
F. J. Lin and J. T. Muckerman, Comput. Phys. Comm. {\bf 63}, 538 (1991)
\bibitem{Schn&Nygaar} B. I. Schneider and Nicolai Nygaard,
J. Phys. Chem. A
{\bf 106}, 10773 (2002).
\bibitem{Res&McC} T. N. Rescigno and C. W. McCurdy, Phys. Rev. A {\bf 62}, 032706-1 (2000)
\bibitem{SchCol&Fed} B. I. Schneider, L. A. Collins and D. L. Feder,
Proceedings of ITAMP Workshop on "Time Dependent Methods For Dynamical
Problems", ed. C. A. Weatherford. To be published in J. Mol. Struc.
\bibitem{VinckeMalegat&Baye} M. Vincke, L. Malegat and D. Baye, J. Phys. B. {\bf 26}, 811 (1993)
\bibitem{Malegat&Vincke} L. Malegat and M. Vincke, J. Phys. B. {\bf 27}, 645 (1994)
\bibitem{Malegat} L. Malegat, J. Phys. B. {\bf 27}, L691 (1994)
\bibitem{Kopal} Z. Kopal, {\em{Numerical Analysis}} (Wiley, 
New York, 1961).
\bibitem{FinEl} J. N. Reddy, {\em An Introduction to the Finite Element Method}, 
(McGraw Hill, New York, 1984)
\bibitem{BessDVR} D. Lemoine, J. Chem. Phys. {\bf 101}, 3936 (1994)
\bibitem{BessProp} G. B. Arfken and H. J. Weber, {\em Mathematical Methods for Physicists},
4$^{th}$ edition, (Academic Press, San Diego, 1995)
\end{thebibliography}
\end{document}